%% file: main.tex
\documentclass[letterpaper,twocolumn,10pt]{article}
\usepackage{deps/usenix-2020-09}

\usepackage{cite}
\usepackage{amsmath,amssymb,amsfonts}
\usepackage{algorithmic}
\usepackage{graphicx}
\usepackage{textcomp}
\usepackage{xcolor}
\def\BibTeX{{\rm B\kern-.05em{\sc i\kern-.025em b}\kern-.08em
    T\kern-.1667em\lower.7ex\hbox{E}\kern-.125emX}}

\input{deps/pkgs}    
\input{deps/macros}  

\begin{document}

\newcolumntype{C}[1]{>{\centering\arraybackslash}p{#1}}

\title{A Comparative Quality Metric for Untargeted Fuzzing with Logic State Coverage}
\author{
  {\rm Gwangmu Lee} \\ {iss300@gmail.com}
}

\thispagestyle{plain}
\pagestyle{plain}

\maketitle


\begin{abstract}

  While fuzzing is widely accepted as an efficient program testing technique, it
  is still unclear how to measure the \emph{comparative quality} of different
  fuzzers. The current de facto quality metrics are edge coverage and the
  number of discovered bugs, but they are frequently discredited by
  inconclusive, exaggerated, or even counter-intuitive results.
  
  To establish a more reliable quality metric, we first note that fuzzing 
  aims to reduce the number of \emph{unknown abnormal} behaviors by observing
  more \emph{interesting} (i.e., relating to unknown abnormal) behaviors. The
  more interesting behaviors a fuzzer has observed, the stronger guarantee it
  can provide about the absence of unknown abnormal behaviors. This suggests
  that the number of observed interesting behaviors must directly indicate the
  fuzzing quality. 
  
  In this work, we propose \emph{logic state coverage} as a proxy metric to
  count observed interesting behaviors. A logic state is a set of satisfied
  branches during one execution, where its coverage is the count of
  \emph{individual} observed logic states during a fuzzing campaign. A logic
  state distinguishes less repetitive (i.e., more interesting) behaviors in a
  finer granularity, making the amount of logic state coverage reliably 
  proportional to the number of observed interesting behaviors.
  We implemented logic state coverage using a bloom filter and performed a
  preliminary evaluation with AFL++ and XMLLint.

\end{abstract}

\section{Introduction}
%

Since AFL \cite{afl} gained popularity, fuzzing has been widely regarded as
one of the most effective software testing methods.
The most basic form is \emph{untargeted} fuzzing, which investigates a whole
program by executing randomly mutated inputs indefinitely without any target
code.
It has shown promising effectiveness on simple software despite its simplicity,  
motivating researchers to further improve it by incorporating various
techniques
\cite{liu2023dsfuzz,deng2023nestfuzz,zheng2023fishfuzz,wu2022strategy,aschermann2019redqueen,rawat2017vuzzer,fioraldi2020aflpp}
and applying it to complex software, such as hypervisors
\cite{liu2023videzzo,myung2022mundofuzz,sergej2021nyx,schumilo2020hypercube} and
language interpreters
\cite{wang2023fuzzjit,wu2023jitfuzz,gross2023fuzzilli,srivastava2021gramatron,chen2021polyglot}.\footnote{In
this paper, we call  \emph{untargeted} fuzzing simply fuzzing if not specified.}

Meanwhile, there have been ever-existing debates on measuring the
\emph{comparative quality} of fuzzing as more research has been done
\cite{wang2020notallcov,wang2019impactcov,bohme2022reliability,klees2018eval,li2021unifuzz}.
While some researchers proposed alternative ways to measure quality
\cite{bohme2021residual,bohme2022reliability,li2021unifuzz}, the
current de facto standard quality metrics are \emph{edge coverage} (i.e., how
many control-flow edges were triggered) and \emph{the number of discovered bugs}
(i.e., how many bugs were discovered). Generally, either larger edge coverage or
more discovered bugs is deemed as a positive indicator of the fuzzing quality.

However, both metrics have been consistently refuted as unreliable. Edge
coverage has long been discredited as a \emph{weak} proxy metric as it
often leads to indecisive measurements; it is common to see the edge coverage
comparisons offering little difference (e.g., equally saturated edge coverage 
\cite{myung2022mundofuzz,sergej2021nyx}) or
influenced by initial inputs (e.g., different requirements for initial inputs 
\cite{chen2021polyglot}).
Furthermore, some experiments yield counter-intuitive results where a code
coverage increment does not translate to more discovered bugs so well
\cite{google-ai}.

The number of discovered bugs, on the other hand, is a much more direct metric
because it counts the very objects (i.e., bugs) that fuzzing aims to reveal.
However, it is highly sensitive to side factors such as bug deduplicatation
method, initial inputs, or the implementation details of both the fuzzer and the
software under test \cite{klees2018evaluating}. Even worse, the number of total
bugs is usually only a handful,
making unclear factors easily introduce a measurement bias that significantly
exaggerates the result in certain software.

In pursuit of a more reliable quality metric for fuzzing, we first note that the
goal of fuzzing, from the software testing perspective, is \emph{minimizing
unknown abnormal program behaviors} (i.e., bugs). Fuzzing approaches this goal
by observing as many \emph{interesting} program behaviors as possible that 
likely reveal abnormal behaviors; the more such behaviors it checks, the fewer
unknown abnormal behaviors are expected to be left.
%
This rationale suggests that \emph{comparing the number of observed interesting
program behaviors between fuzzers} must indicate its \emph{comparative
quality}, as a larger number offers a greater guarantee of fewer unknown
abnormal behaviors. 

As it is challenging to directly count interesting program behaviors when
fuzzing produces hundreds of them every second, we propose \emph{logic state
coverage} as a feasible proxy metric of interesting behaviors. A logic state is
a set of all satisfied branches during one execution, encoding the
exhibited program behavior without branch repetition information.
As a consequence, less repetitive (i.e., more \emph{interesting}) program
behaviors are reflected as more \emph{distinct} logic states. Logic state
coverage counts all distinct observed logic states during a fuzzing campaign,
which means that a larger coverage suggests more observed interesting behaviors.



\section{Formalizing Fuzzing}

In this section, we formalize the fuzzing procedure to derive a strong quality
metric. We first start with defining a program
behavior (\autoref{s:fuzz:behave}) and, based on the definition, formalize the
procedure of untargeted fuzzing (\autoref{s:fuzz:unfuzz}).

\pp{Notation and assumption}
We denote a set and its element in bold and lowercase, respectively.  For
example, $\mathbb{I}$ and $\mathbb{O}$ denote a set of all inputs and outputs
respectively, where $i$ and $o$ are the element of each set (i.e., an input and
an output). If not specified, all arguments assume single-thread programs.

\subsection{Program Behavior}
\label{s:fuzz:behave}


A program $p$ is a map that cast an input $i$ to an observable output $o$ and
a \emph{program behavior} $h$, which can be defined as a stream of executed
instructions. Formally, if $\mathbb{H}$ is a \emph{program behavior space}, 
a set of all possible program behaviors,
\begin{equation}
  p(i) = (o, h) \iff p: \mathbb{I} \rightarrow \mathbb{O} \times \mathbb{H}.
\end{equation}

\pp{Program conditions}
Program conditions are the instructions that affect the control flow and
transform the program behavior accordingly. They are further
broken down to two kinds: branch and exceptional.
Branch conditions (or \emph{branches}) decide the regular control
flow between basic blocks. Exceptional conditions (or \emph{exceptions}), on the
other hand, interrupt the control flow and abnormally terminate the
program.\footnote{C++-style exceptions, despite its name, can be regarded as a
form of branch conditions as they transfer the control flow to another basic
block.} 

\pp{Normality of program behaviors}
Program behaviors can be divided into two classes depending on the satisfied
program conditions: normal and abnormal. Normal program behaviors only satisfy
regular branches and terminate at program exit points. Abnormal
program behaviors, on the other hand, satisfy at least one exception and
terminate at non-exit points. Sanitization
\cite{yun2016apisan,serebryany2012asan,kasan,ktsan,kubsan,stepanov2015msan,ubsan} can be seen as bringing
\emph{unintended} behaviors (e.g., integer-overflow or use-after-free) to the
\emph{abnormal} territory by adding exceptions.

\pp{Uniqueness of program behaviors}
For a single-thread program, 
program behaviors are exclusively determined by the history of program
conditions. This is because a single-thread program always executes the 
instructions sequentially until the next program condition, making every
portion of the instruction stream \emph{implied} by terminating program
conditions.

Formally, let $\mathcal{T}_b$ and $\mathcal{T}_e$ be the history of branches and
exceptions. Then, for a single-thread program, there exists a function
$f(\mathcal{T}_b, \mathcal{T}_e) = h$ that uniquely maps a pair
$(\mathcal{T}_b, \mathcal{T}_e)$ to a certain program behavior $h$.
%


\subsection{Untargeted Fuzzing}
\label{s:fuzz:unfuzz}

Fuzzing, as a form of software testing, aims at minimizing the number of unknown
abnormal program behaviors. \emph{Untargeted} fuzzing addresses this goal by:
(i) investigating as many \emph{new} program behaviors as possible and (ii)
keeping the mutation base input \emph{interesting} (i.e., more likely revealing
unknown abnormal behaviors after mutation).

\pp{New vs. unknown behaviors}
We first make a clear distinction between \emph{new} and \emph{unknown} 
behaviors. \emph{New} behaviors are, as the name suggests, simply the behaviors 
that have not been observed until now. On the other hand, \emph{unknown}
behaviors are a subset of the \emph{new} behaviors that are \emph{not
equivalent} to any observed behaviors. In particular, an abnormal behavior is
\emph{unknown} if it is not equivalent to any observed abnormal behaviors (e.g.,
a unique root cause). Formally, if $\mathrm{U}$ and $\mathrm{N}$ are the sets of
\underline{\textbf{u}}nknown and \underline{\textbf{n}}ew behaviors,
$\mathrm{U} \subset \mathrm{N}$.

\pp{Formal justification}
The approach of untargeted fuzzing can be formally described as follows. Upon
executing a mutated input, let $\Pr(\mathrm{N})$ be the probability of observing a
\underline{\textbf{n}}ew behavior as a result of execution. Then, the
probability of discovering \underline{\textbf{u}}nknown
\underline{\textbf{a}}bnormal behavior $\Pr(\mathrm{UA}) := \Pr(\mathrm{U} \cap
\mathrm{A})$ is
\begin{align}
  \Pr(\mathrm{UA}) 
      &= \Pr(\mathrm{N} \cap \mathrm{UA}) \tag{by $\mathrm{U} \subset \mathrm{N}$} \\
      &= \Pr(\mathrm{N}) \cdot \Pr(\mathrm{UA} \mid \mathrm{N}) \tag{by chain rule}. 
\end{align}

Here, the second term $\Pr(\mathrm{UA} \mid \mathrm{N})$ is the expectation of
revealing an unknown abnormal behavior given a new behavior, which can be
interpreted as the \emph{interestingness} of the mutation base. In other words,
untargeted fuzzing attempts to (i) increase $\Pr(\mathrm{N})$ to investigate
more program behaviors while (ii) keeping the interestingness $\Pr(\mathrm{UA}
\mid \mathrm{N})$ high.
%

\pp{Interestingness in practice}
As the interestingness $\Pr(\mathrm{UA} \mid \mathrm{N})$ is the expectation of
an \emph{unobserved} behavior, there is no direct way to measure
interestingness. 
However, the common usage of corpus minimization
\cite{libfuzzer,efffuzz,herrera2021seedsel,herrera2021seedsel} and branch hit
count buckets \cite{afl,libfuzzer} in untargeted fuzzing suggests that a program behavior
is less likely to be \emph{unknown abnormal} (i.e., less interesting) if its
mutation base input is highly repetitive.
%
The rationales are: (i) if a fewer repetition was normal,
more repetition is also likely normal and (ii) even if it was abnormal, it is 
not unknown if it corresponds to pre-observed less-repetition counterparts. 

To generalize this practice from mutation base inputs to all inputs (i.e., all
their program behaviors), a program behavior can be deemed less interesting if
it include more repetitions.

%
%




\pp{Quality of untargeted fuzzing}
As the procedure of untargeted fuzzing can be seen as maximizing observed
interesting behaviors in the program behavior space, the portion of such 
behaviors in the space directly indicates the \emph{quality} of the
procedure: the more it checks such behaviors, the less the unknown
abnormal behaviors are expected to be left.

Based on this, we can define the \emph{comparative quality} between
different untargeted fuzzing techniques as follows: given the same amount of
time and computing resource, \textbf{how many more interesting program behaviors
can it observe than others?} We argue that this is a \emph{strong quality
metric} for untargeted fuzzing as it directly compares the reduced expectation
of unknown abnormal behaviors. 


\section{Logic State Coverage}
\label{s:lscov}

In theory, interesting behaviors can be naively counted by investigating
every program behavior. However, this naive approach is infeasible as
fuzzing produces hundreds of arbitrarily long instruction streams \emph{every
second}.
In this section, we propose \emph{logic state coverage} as an indirean indirectt proxy of
counting interesting behaviors.

\begin{figure}[t]
  \centering
  \includegraphics[width=\columnwidth]{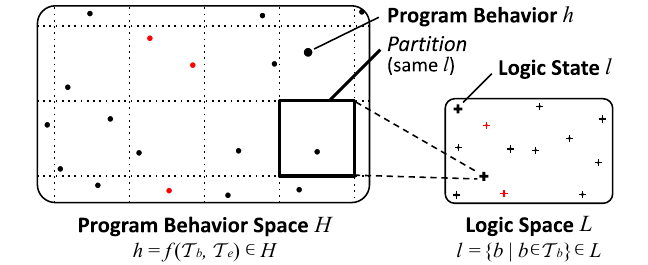}
  \caption{The program behavior space and the logic space. Red indicates
  abnormal program behaviors or logic states}.
  \label{f:prog-to-logic}
\end{figure}

\pp{Logic state and logic space}
A \emph{logic state} $l$ is a set of all satisfied branches during execution,
which is essentially the order-ignored version of a branch history. Formally,
given a branch history $\mathcal{T}_b$ of an execution,
\begin{equation}
\label{e:ls}
  l := \{b \mid b \in \mathcal{T}_b\}.
\end{equation}

A \emph{logic space} $\mathbb{L}$, then, is a set of all possible logic states
of a given program. 
It is possible to see a logic space as the \emph{subdivision} of a program
behavior space along logic states. \autoref{f:prog-to-logic} shows the
relationship between a program behavior space and a logic space, where each
\emph{partition} of a subdivided program behavior space corresponds to one logic
state.

\pp{Logic state coverage}
\emph{Logic state coverage} is \emph{the count of observed distinctive logic
states during fuzzing}, where a logic state is called \emph{covered} when
fuzzing observed a program behavior that belongs to such a logic state (i.e.,
triggering all and only the branches in the logic state). 
Analogically, logic state coverage represents the \emph{covered area} of a logic
space.

Logic state coverage is crucially distinguished from edge coverage in that, 
while edge coverage merges all observed edge traces into one, logic state
coverage distinguishes every logic state and counts them individually. This
enables logic state coverage to count underlying program behaviors independent 
of what has been observed before, while edge coverage unpredictably ignores some
of them as it is heavily affected by the previous observation.

As shown in \autoref{f:prog-to-logic}, a logic state (i.e., a partition in the
program behavior space) may contain various program behaviors, suggesting a
single covered logic state may encompass several different behaviors. However,
in \autoref{s:prop}, we explain that a logic state is representative of
interesting program behaviors as it (i) reflects the interestingness of
underlying program behaviors and (ii) only contains the program behaviors of the
same normality.

\section{Representativeness of Logic States}
\label{s:prop}

In this section, we discuss the desirable properties of logic states to
represent interesting program behaviors (\autoref{s:prop:prop}). Then, we
continue to discuss how logic states have such properties
(\autoref{s:prop:repr} and \ref{s:prop:local}).

\subsection{Desirable Properties}
\label{s:prop:prop}

\pp{Interestingness}
Logic states should be sufficiently representative of the \emph{interesting}
program behaviors. Specifically, it should distinguish most program behaviors
deemed interesting while not over-representing less interesting ones. 

\pp{Normality}
The \emph{normality} of all program behaviors should be uniform (i.e., the same)
within a logic state. Otherwise, the normality of one observed program behavior
cannot represent that of other behaviors in the same logic state.
%

\begin{figure}[t]
  \centering
  \vspace{-1em}
  \includegraphics[height=4.5cm]{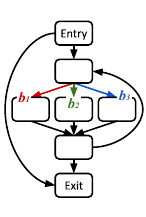}
  \includegraphics[height=4.5cm]{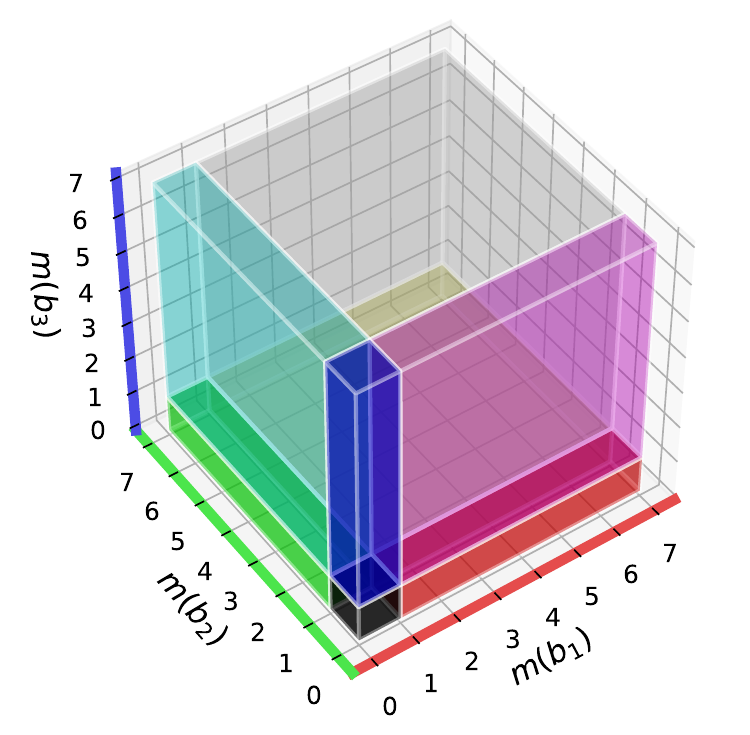}
  \vspace{-.5em}
  \caption{Example control flow graph and its logic states (colored box) that
  the program behaviors with different branch hit counts belong to. $m(b)$
  denotes the hit count of branch $b$.}
  \label{f:repr}
\end{figure}

\subsection{Interestingness}
\label{s:prop:repr}

Logic states coalesce less interesting program behaviors that exhibit more
branch repetitions in a fewer logic states, resulting in smaller (larger) logic
state coverage as it observes more uninteresting (interesting) program
behaviors.


\autoref{f:repr} shows an example control-flow graph and its logic space,
gridded up by the different hit counts $m(b)$ of the branch $b_1$, $b_2$, and $b_3$.
Each colored box represents one logic state, which includes all program
behaviors that fall into the box.
For example, a program behavior that does not hit any of $b_1$, $b_2$, and $b_3$ 
(i.e., $m(b_1)=m(b_2)=m(b_3)=0$) belongs to the \cbox{black} logic state, while
all program behaviors that hit only $b_1$ (i.e., $m(b_1)>1$ and
$m(b_2)=m(b_3)=0$) fall into \cbox{red}.

The colored boxes in \autoref{f:repr} suggest the increasing size of logic
states as their constituent program behaviors involve more repetitive branches.
To be specific, a logic state covers a progressively larger volume 
in the logic space 
(i.e., \cbox{black} $\rightarrow$ \cbox{red}/\cbox{green}/\cbox{blue}
$\rightarrow$ \cbox{yellow}/\cbox{cyan}/\cbox{magenta} $\rightarrow$
\cbox{gray}) as it contains more repetitive branches (i.e., 0 $\rightarrow$ 1
$\rightarrow$ 2 $\rightarrow$ 3 repetitive branches). 
This results in repetitive (i.e., uninteresting) program states packed
to fewer logic states, suppressing (boosting) logic state coverage as it observes more
uninteresting (interesting) behaviors.


\begin{figure}[t]
  \centering
  \includegraphics[width=\columnwidth]{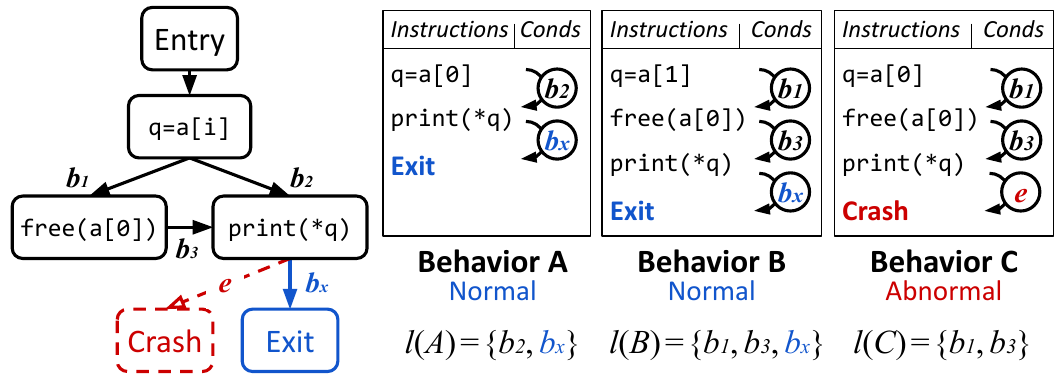}
  \caption{Example control flow graph and program behaviors. The exceptional
  condition $e$ is satisfied when the memory pointed by $q$ is a freed
  memory. $l(X)$ denotes the logic state of a program behavior $X$.}
  \label{f:local}
\end{figure}

\subsection{Normality}
\label{s:prop:local}

Given that a program does not exhibit an exception exactly at the program exit
point, all program behaviors in one logic state have the same normality, namely
either normal or abnormal altogether.
This is because normal behaviors trigger the \emph{exit branches} that
lead to the exit point, while abnormal behaviors do not. Since the logic
states of two behaviors are always distinguished by exit branches, they cannot
have the same set of satisfied branches (i.e., the same logic state).

\autoref{f:local} shows an example program that contains a use-after-free
exception $e$, which is satisfied when the memory pointed by $q$ belongs to
freed memories. Three example program behaviors and their logic states are shown
on the side. Since normal behaviors (Behaviors A and B) terminate at the program
exit, their logic states contain the exit branch $b_x$ that directly leads to
the program exit. On the other hand, since the abnormal behavior (Behavior C)
triggers the exception $e$ that crashes the execution before reaching the
program exit, its logic state does not contain any exit branch unlike normal
behaviors.

\section{Measuring Logic State Coverage}
\label{s:measure}

In this section, we first summarize the high-level procedure of logic space
coverage measurement and its subsequent requirements 
(\autoref{s:measure:req}). Then, we introduce a \emph{bloom filter} as an
effective way to achieve the requirements (\autoref{s:measure:bloom}).

\subsection{High-level Procedure}
\label{s:measure:req}


%
At a high level, the measurement can be described as follows:

\begin{enumerate}[noitemsep,label=\arabic*)]
  \item \emph{Record} a logic state per execution.
  \item \emph{Add} the logic state to a set of observed logic states.
  \item \emph{Count} the size of the set, yielding logic state coverage.
\end{enumerate}

In this procedure, recording a logic state (1) is technically identical to edge
tracing in coverage-guided fuzzers \cite{afl,fioraldi2020aflpp,libfuzzer}.
Managing logic state coverage (2 and 3), on the other hand, poses an extra
challenge as fuzzers are expected to observe a myriad of distinct logic states
during a campaign.

\pp{Requirements}
As the purpose of logic state coverage is comparing (i.e., evaluating) different
fuzzers, the measurement should not induce any bias to certain fuzzers. To this
end, it should satisfy the following requirements.

\begin{itemize}

  \item \textbf{R1. Light computation overheads.}
    The computation overheads should not be heavy. Otherwise, it would overshadow
    the benefit of fast fuzzing executions, penalizing high-speed fuzzer designs
    \cite{sergej2021nyx,schumilo2020hypercube,song2020agamotto}.

  \item \textbf{R2. Constant computation overheads.}
    The measurement overheads should not increase as logic state coverage
    expands. Otherwise, it would penalize the fuzzers with larger logic space
    coverage.

  \item \textbf{R3. Feasible memory overheads.}
    The memory requirement for the measurement should be feasible for commodity
    evaluation settings for practical reasons.
\end{itemize}

\subsection{Leveraging a Bloom Filter}
\label{s:measure:bloom}

\begin{figure}[t]
  \centering
  \includegraphics[width=\columnwidth]{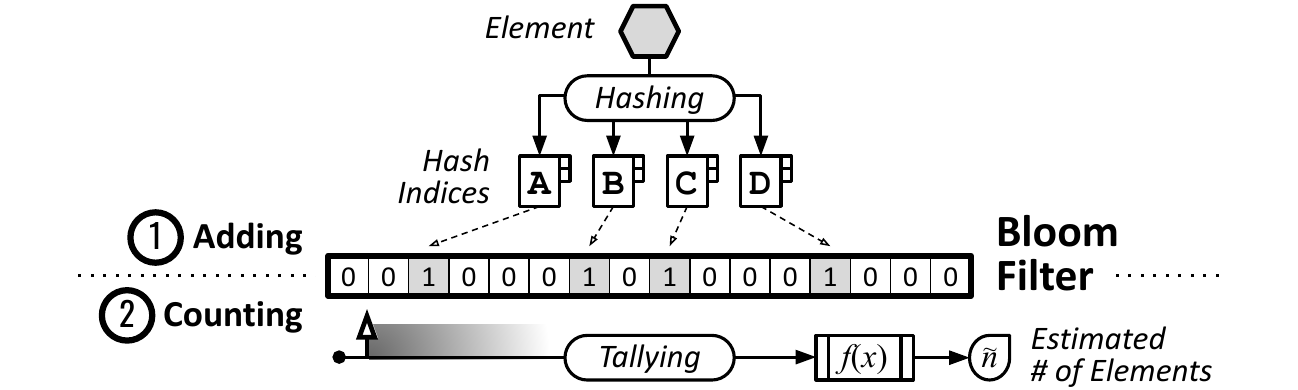}
  \caption{Operation illustration of a bloom filter.}
  \label{f:bloom}
\end{figure}


For logic state coverage management, we employ a \emph{bloom filter} that 
offers an 
efficient set implementation at the cost of manageable inaccuracy (i.e., false
positives).
%
%
\autoref{f:bloom} illustrates the basic operations of a bloom filter. Starting
from an array of 0s, it \emph{adds} an element by setting the hash indices to 1
and \emph{counts} the number of elements by tallying 1s in the filter and
applying it to the counting formula (\autoref{e:count}).

%
\pp{Usage and benefit}
In the measurement, a bloom filter represents \emph{a set of observed logic
states}, which accepts a logic state every execution and yields logic state
coverage via the number of elements in the set. It only requires a few
computations for hashes and the counting formula (\textbf{R1}), both of which
are nearly $O(1)$ as the number of hashes and the filter size are fixed
(\textbf{R2}). Furthermore, the filter size is entirely agnostic to the size of
a logic state thanks to hashing (\textbf{R3}). 

\pp{Deciding parameters}
A bloom filter may introduce some inaccuracy by construction. However, it can
be reliably suppressed by the choice of data structure parameters, namely the
filter size (i.e., the number of bits) $n_b$ and the number of hashes $n_h$.
They can be optimally determined by the maximum expected number of elements
$n_e$ and the desired false positive probability $\epsilon$ by the following
equations \cite{tarkoma2012bloom}:
\begin{equation}
  n_b = -n_e\ln{\epsilon}/(\ln{2})^2, \enspace n_h = -\log_2{\epsilon},
  \label{e:param}
\end{equation}

In our setting, $n_e$ corresponds to the upper bound of \emph{observable
distinct logic states} during a fuzzing campaign. To conservatively estimate
$n_e$, we adopt three premises in ideal evaluation : (i) fuzzing continues for
the maximum 24 hours, (ii) a fuzzer executes the maximum 1,000 inputs per
second, and (iii) \emph{every} execution yields a distinctive logic state. Then, 
\begin{align}
  n_e &= \SI{24}{\hour} \times \SI[per-mode=symbol]{3600}{\second\per\hour}
         \times \SI[per-mode=symbol]{1000}{\exec\per\second} 
         \times \SI[per-mode=symbol]{1}{\state\per\exec} \nonumber \\
      &= \SI{86.4e+6}{\state} 
         \approx \SI[parse-numbers=false]{84 \times 2^{20}}{\state}. \nonumber
\end{align}

Assigning the upper bound $n_e$ and a reasonable false positive probability
$\epsilon=0.05$ to \autoref{e:param} yields:
\begin{equation}
  n_b = \SI{538e+6}{\bit}\approx\,\SI{512}{\mbit}=\SI{64}{\mb}, \enspace n_h = 4.32 \approx 4. \nonumber
\end{equation}

Notice that adding a logic state only requires calculating four hashes per
execution (\textbf{R1} and \textbf{R2}) and the memory overhead ($\SI{64}{\mb})$
is also marginal to a commodity setup (\textbf{R3}).

\pp{Counting elements}
Along with data structure parameters, a counting formula that estimates the
number of elements may also affect the accuracy of logic state coverage.
Papapetrou et al. \cite{papapetrou2010cardinality} proposed a high-accuracy
counting formula that calculates the estimated number of elements
$\tilde{n}_e(X)$ when the number of 1s in the filter is $X$ as follows:
\begin{equation}
  \tilde{n}_e(X) = \frac{\ln{(1-X/n_b)}}{n_h \ln{(1-1/n_b)}}.
  \label{e:count}
\end{equation}

\autoref{e:count} comes with three major benefits. First, it offers one of the
best accuracy among all alternatives (c.f., only 4\% error even in an extreme
filter density of 90\%)
\cite{harmouch2017cardinality,papapetrou2010cardinality}. 
%
Second, it allows for a near-constant-time calculation as it only requires the
number of 1s apart from the formula calculation, as well as empirically shown in
\cite{harmouch2017cardinality} (\textbf{R2}). 
Finally, it presents a calculable error bound that the actual number of elements
likely falls within given a confidence probability \cite{papapetrou2010cardinality}.


\section{Design and Implementation}
\label{s:design}

In this section, we first present the design overview (\autoref{s:design:over})
and elaborate on details (\autoref{s:design:rec} and
\ref{s:design:count}). The implementation in progress can be found at
\smaller[0.5]\url{https://github.com/gwangmu/LogicStateCoverage}.

\begin{figure}[t]
  \centering
  \includegraphics[width=\columnwidth]{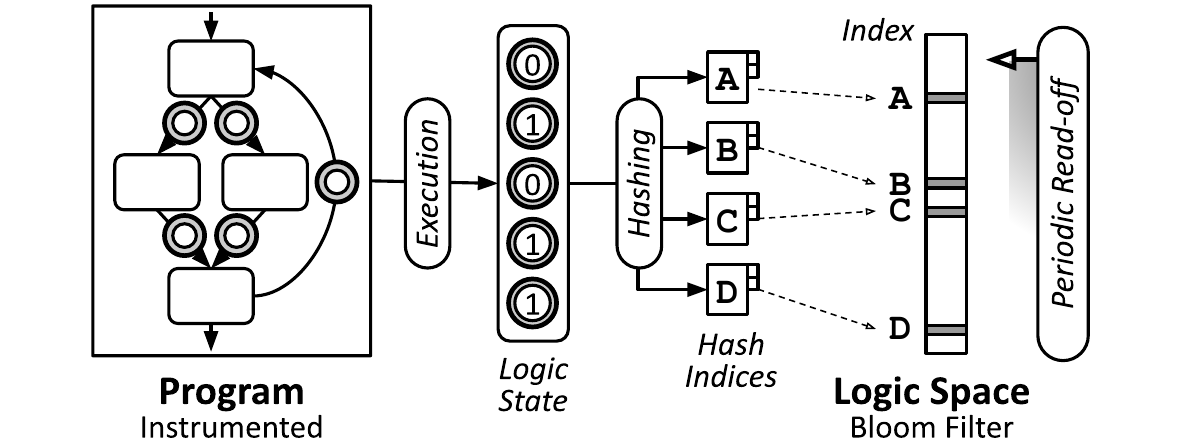}
  \caption{Logic state coverage measurement overview.}
  \label{f:design}
\end{figure}

\subsection{Overview}
\label{s:design:over}


\autoref{f:design} illustrates the workflow of logic state coverage
measurement. Similar to edge coverage, it first records the \emph{logic state}
of each execution via instrumentation (\autoref{s:design:rec}). Then, it adds
the recorded logic state to a \emph{bloom filter} that represents a set of
observed logic states, by setting the hash indices (\autoref{s:design:up}).
Finally, the measurement periodically reads off the bloom filter to count the
number of observed logic states (\autoref{s:design:count}).

\subsection{Recording a Logic State}
\label{s:design:rec}


Recording a logic state is equivalent to tracing executed edges in conventional
edge coverage. Specifically, a taken edge is calculated by combining the
hashes of two consecutively executed basic blocks, which is then added to the
logic state of the current execution. Notice that a logic state is \textit{per
execution} and is not merged throughout multiple executions as in edge
coverage. We implemented logic state recoding based on the LLVM pass for
\cc{afl-clang-fast} \cite{afl}.

\subsection{Updating Logic State Coverage}
\label{s:design:up}


The recorded logic state is updated to the bloom filter at the end of every
execution. In particular, a number of hashes are calculated from the entire
logic state (i.e., \cc{A}, \cc{B}, \cc{C}, and \cc{D} in \autoref{f:design}) and
are used as bloom filter indices to be set. We used MurMurHash3
\cite{murmurhash3} to hash a logic state.

\begin{figure*}[t]
  \centering
  \includegraphics[width=0.8\textwidth]{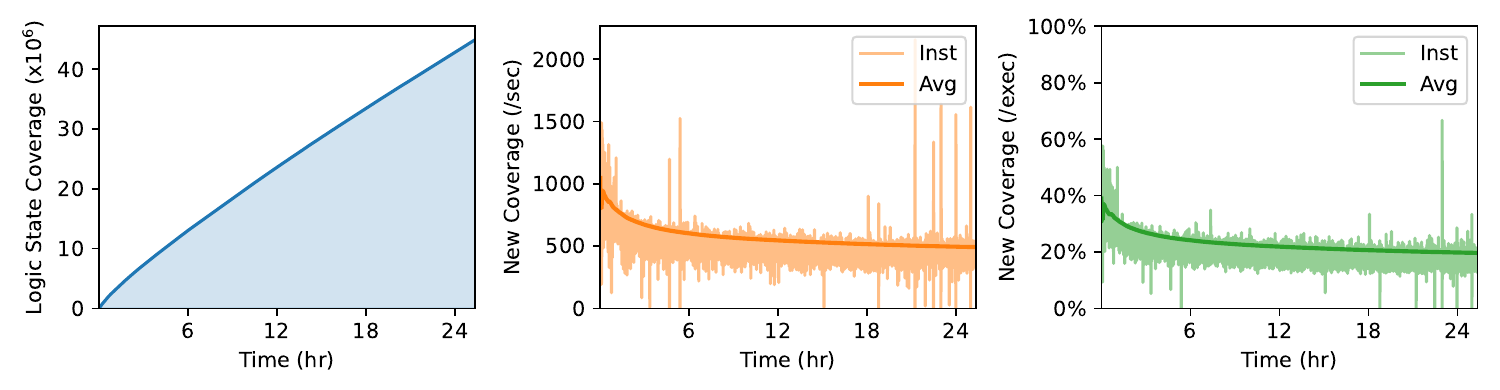}
  \caption{24-hour fuzzing campaign of XMLLint on AFL++. (Ins: instantaneous,
    Avg: average)}
  \label{f:aflpp-xmllint-epfl}
\end{figure*}

\subsection{Counting Logic State Coverage}
\label{s:design:count}

Finally, the logic state coverage of the fuzzing campaign is measured by
periodically reading off the bloom filter, where the number of 1s in the filter
is fed to \autoref{e:count} to count the distinctive number of observed logic
states. We implemented logic state coverage measurement as a separate process,
which a measurer should launch before starting a fuzzing campaign.

\section{Evaluation}

\subsection{Preliminary Result}

\autoref{f:aflpp-xmllint-epfl} shows the preliminary evaluation result
using XMLLint and AFL++. The evaluation was done on an Intel Core i7-8665U CPU
machine with 16 GB of memory. The logic state coverage (left) shows the
distinctive logic states covered over time. The new coverage per second (middle)
shows the number of newly discovered logic states per time, while the new
coverage per execution (right) shows the percentage of executions that
discovered new logic states per execution. \textsc{Avg} indicates the average
new coverage per second or per execution, and \textsc{Ins} indicates the
instantaneous new coverage during the last 10-second measurement frame.

The logic state coverage increases steadily over time but gradually slows
down toward the 24-hour mark. This trend is also reflected in the new coverage
plot, where the average new coverage slowly decreases to a steady point toward
the end of the fuzzing campaign.
Notice that the sporadic spikes in the instantaneous new coverage (\textsc{Ins})
are more in the per-second plot than in the per-execution plot, suggesting that
these sudden new coverage surges were primarily caused by quicker execution. 

\subsection{Research Questions}

Some of the prime research questions that further evaluation should answer
include the following.

\begin{itemize}
  \item \textbf{RQ1} Can logic state coverage be implemented with all
    requirements in \autoref{s:measure:req} satisfied?
  \item \textbf{RQ2} How does logic state coverage measure the comparative
    quality of popular untargeted fuzzers?
  \item \textbf{RQ3} Can logic state coverage distinguish the fuzzers
    indistinguishable from conventional edge coverage?
\end{itemize}

For \textbf{RQ1}, the evaluation may evaluate the computation and memory
overhead added by the logic state coverage measurement. It should not perturb
the execution speed of measured fuzzers to minimize the measurement bias. For
\textbf{RQ2}, the evaluation may compare the logic state coverage of popular
untargeted fuzzers \cite{afl,fioraldi2020aflpp,libfuzzer,honggfuzz} with a
set of standard fuzzing targets (e.g., FuzzBench \cite{metzman2021fuzzbench}).
For \textbf{RQ3}, the evaluation may compare the logic state coverage of
specialized fuzzers (e.g., hypervisor fuzzers
\cite{liu2023videzzo,myung2022mundofuzz,sergej2021nyx,schumilo2020hypercube} and
grammar fuzzers
\cite{wang2023fuzzjit,wu2023jitfuzz,gross2023fuzzilli,srivastava2021gramatron,chen2021polyglot}),
whose quality has not been well-distinguished by a conventional edge coverage
metric.


\section{Discussion}
\label{s:disc}

\pp{Completeness of untargeted fuzzing}
Unlike comparative quality that only requires to compare the number of observed
distinct program behaviors, it is difficult to measure the absolute
\emph{completeness} of untargeted fuzzing that requires to measure how much a
program has been completely verified. This is because the total number of
program behaviors is unknown: if it can be known, then we would already know
which abnormal behaviors are possible, precluding the necessity of fuzzing in
the first place.

One possible compromise is performing preliminary static analysis to
\emph{estimate} the total number of program behaviors, including abnormal ones.
However, it would pose a potential problem that the estimated 100\% completeness
does not necessarily mean the actual 100\% completeness, defeating the very
purpose of measuring \textit{completeness}.

\pp{Generalization to multi-thread programs}
Unlike single-thread programs, \emph{multi-thread} programs may concurrently
execute multiple basic blocks. In this case, a program behavior (i.e., a stream
of executed instructions) also depends on the thread interleaving in
addition to the branch and exception histories, which breaks the proxy
properties of a logic space (i.e., interestingness and normality).
An alternative proxy of comparative quality may be proposed by either
extending logic state coverage or devising yet another coverage metric.

\pp{Normality of a partial logic state}
It is crucial to notice that an abnormal behavior \emph{still follows the
regular control flow up until any exception happens}. In \autoref{f:local}, the
abnormal behavior (Behavior C) follows the same control flow to a normal
behavior (Behavior B) until use-after-free, which makes the behavior eventually
abnormal unlike its normal counterpart. 
What it implies is that the executions that do not contribute to the growth of
logic state coverage are not entirely wasteful executions because they are still
possible to reveal abnormal behaviors. However, they are unlikely to do so, let
alone revealing \emph{unknown} ones (\autoref{s:fuzz:unfuzz}).

\pp{Quality metric for general fuzzing}
Logic state coverage evaluates the quality of \emph{untargeted} fuzzing, where
the program behaviors with less repetition are generally deemed more
interesting. 
However, \emph{targeted} fuzzing
\cite{aflchurn,chen2020savior,osterlund2020parmesan,lee2024syzrisk} attempts to
capitalize the recent findings \cite{aflchurn,zhai2022ndss,nikolaos2022longvuln}
that some code such as recent patches are potentially \emph{more interesting}
(i.e., likelier to cause abnormal behaviors), which may render logic state
coverage to under-estimate targeted fuzzing by overlooking potentially
interesting behaviors.

To devise a quality metric that also encompasses targeted fuzing, future
research may need to address two issues revolving around interestingness. First,
it should accurately describe the interestingness in an impartial way (i.e., not
favoring a certain fuzzer or setting). To do so, it should assess which factor
affects interestingness and how much more than one another. Second, it should
incorporate such a description into the quality metric. Should it follow the
philosophy of logic state coverage, grouping uninteresting behaviors may be a
possible approach to make them less significant.

\pp{Alternative ways to manage logic state coverage}
There has been much research work regarding how to estimate the cardinality
(i.e., the number of elements) in a set in an efficient way
\cite{papapetrou2010cardinality,harmouch2017cardinality}, where using a bloom
filter is just one of the approaches. Harmouch et al.
\cite{harmouch2017cardinality} evaluated different cardinality approximation
approaches and compared their accuracy with various benchmarks.  Although using
a bloom filter is the most recent approach compared in the paper, other
alternative approaches can be incorporated to manage logic state coverage as it
is not always the optimum solution.

\pp{Semantically abnormal behaviors}
While the definition of program behaviors is separated from the output, it can
still incorporates \emph{semantically} abnormal behaviors (i.e., semantic bugs)
by means of exceptions. To be specific, semantic error detection such as
semantic sanitizers \cite{kim2019hydra,yun2016apisan} may capture semantic bugs
on the fly and make the behavior abnormal by terminating it. The definition
allows this because the output is just an observable subset of the data context
during the execution, which exceptions can reference at runtime.

\section{Related Work}

\pp{Reliability of conventional metrics}
Since the advent of popular fuzzing, researchers have consistently
called into question a method to evaluate fuzzers. Klees et al.
\cite{klees2018evaluating} evaluated then state-of-the-art fuzzers and raised an
issue about using edge coverage and the number of discovered bugs as quality
metrics. \textsc{UniFuzz} \cite{li2021unifuzz} also pointed out that such
metrics are incomprehensible and lead to incomplete assessment. B\"{o}hme et al.
\cite{bohme2022reliability} also noted that edge coverage does not strongly
represent the comparative quality between fuzzers.  While they all suggest the
necessity of an alternative quality metric, it has yet to be proposed and widely
accepted in practice. 

\pp{Alternative metrics}
Some researchers did propose alternative metrics for coverage-guided fuzzing
\cite{wang2020notallcov,wang2019impactcov,yan2020pathafl,gan2020greyone,gan2018collafl}.
However, they are commonly meant for \emph{performance} (e.g., more bugs or edge
coverage), not for \emph{comparison}, which makes using such metrics for
comparison likely yield an unjustified advantage toward certain fuzzers.
\textsc{Sand} \cite{kong2024sand} proposed \emph{execution patterns}
(conceptually similar to logic states) to enable fuzzing without sanitization.
However, similar to other alternative metrics, it was never meant for
comparative evaluation nor correlated to the fuzzing quality. 

\pp{Evaluation methods and platforms}
%
B\"{o}hme et al. \cite{bohme2021residual} proposed a technique to estimate the
probability of discovering more bugs,
but the probability cannot be directly compared between fuzzers as it highly
depends on the capability of individual fuzzers.
%
\textsc{UniFuzz} \cite{li2021unifuzz} and \textsc{FuzzBench}
\cite{metzman2021fuzzbench} presents a platform to comparatively evaluate
fuzzers in multiple metrics. 
%
Logic state coverage can be readily incorporated into the platform as a quality
metric.

\section{Conclusion}

While fuzzing is widely accepted as an efficient program testing technique, 
it still lacks the metric to comparatively measure the quality of various
designs and implementations. To establish a reliable quality metric, we
note that the more interesting behaviors a fuzzer has observed, the stronger
guarantee it can provide about the absence of unknown abnormal behaviors. Based
on this, we propose \emph{logic state coverage} as a proxy metric to count
observed interesting behaviors. A logic state is a set of satisfied branches
during one execution, where its coverage is the count of \emph{individual}
observed logic states during a fuzzing campaign. We implemented logic state
coverage using a bloom filter and performed a preliminary evaluation with AFL++
and XMLLint.



\input{main.bbl}

\begin{appendix}

\end{appendix}

\end{document}

%% file: deps/pkgs.tex
\usepackage{acronym}
\usepackage{xspace}
\usepackage{xstring}
\usepackage{tikz}
\usepackage{hyperref}
\usepackage{censor}
\usepackage{multirow}
\usepackage{booktabs}
\usepackage{makecell}
\usepackage{listings}
\usepackage{relsize}
\usepackage{xcolor}
\usepackage{nimbusmono}
\usepackage[T1]{fontenc}
\usepackage{underscore}
\usepackage{tcolorbox}
\usepackage{enumitem}
\usepackage[explicit]{titlesec}
\usepackage{subcaption}
\usepackage{lipsum}
\usepackage{tabularx}
\usepackage{graphicx}
\usepackage{siunitx}
\usetikzlibrary{arrows.meta}

%% file: deps/macros.tex
\newcounter{gm} 

\ifx\enablecomm\undefined
  \newcommand{\gm}[1]{}
\else
  \newcommand{\gm}[1]{\textcolor{teal}{\{Gwangmu-\arabic{gm}: #1\}}\addtocounter{gm}{1}}
\fi

\newcommand{\ignore}[1]{}

\newcommand{\cc}[1]{\mbox{\smaller[0.5]\texttt{#1}}}

\newcommand{\pp}[1]{\vspace{2px}\noindent{\bf \IfEndWith{#1}{.}{#1}{#1.}}}
\newcommand{\ppi}[1]{\vspace{10px}\noindent{\it \hspace{6px}\IfEndWith{#1}{.}{#1}{#1.}}\vspace{10px}}

\definecolor{DGreen}{rgb}{0,0.4,0.05}
\definecolor{DBlue}{rgb}{0.02, 0.2, 0.5}
\definecolor{DRed}{rgb}{0.7, 0.2, 0.1}
\lstdefinelanguage{diff}{
  morecomment=[f][\color{DBlue}]{@@},     
  morecomment=[f][\color{DRed}]-,         
  morecomment=[f][\color{DGreen}]+,       
  morecomment=[f][\color{purple}]{---}, 
  morecomment=[f][\color{purple}]{+++},
}

\makeatletter
\newcommand\notsotiny{\@setfontsize\notsotiny{6.2}{7}}
\makeatother

\definecolor{DBack}{rgb}{0.98, 0.98, 0.98}
\definecolor{DFrame}{rgb}{0.5, 0.5, 0.5}
\lstdefinestyle{diff}{
  numbers=left, 
  basicstyle=\ttfamily\notsotiny,
  xleftmargin=1.5em,
  frame=l,
  framexleftmargin=0.3em,
  frame=single,
  captionpos=b,
  language=diff,
  backgroundcolor=\color{DBack},
  rulecolor=\color{DFrame},
  escapeinside={<@}{@>}
}

\lstdefinestyle{c}{
  numbers=left, 
  basicstyle=\ttfamily\scriptsize,
  xleftmargin=1.5em,
  frame=l,
  framexleftmargin=0.3em,
  frame=single,
  captionpos=b,
  language=c,
  escapeinside={<@}{@>},
  keywordstyle=[2]{\texttt},
  backgroundcolor=\color{DBack},
  rulecolor=\color{DFrame},
  morekeywords=[2]{type}
}

\lstdefinestyle{python}{
  numbers=left, 
  basicstyle=\ttfamily\scriptsize,
  xleftmargin=1.5em,
  frame=l,
  framexleftmargin=0.3em,
  frame=single,
  captionpos=b,
  language=Python,
  escapeinside={<@}{@>},
  keywordstyle=[2]{\texttt},
  backgroundcolor=\color{DBack},
  rulecolor=\color{DFrame},
  morekeywords=[2]{type}
}

\lstdefinestyle{scala}{
  numbers=left, 
  basicstyle=\ttfamily\scriptsize,
  xleftmargin=1.5em,
  frame=l,
  framexleftmargin=0.3em,
  frame=single,
  captionpos=b,
  escapeinside={<@}{@>},
  backgroundcolor=\color{DBack},
  rulecolor=\color{DFrame},
  keywordstyle=[2]{\textbf},
  morekeywords = [2]{def},
  morekeywords = [2]{val},
  morekeywords = [2]{Boolean},
  morekeywords = [2]{return}
}

\DeclareSIUnit{\exec}{exec}
\DeclareSIUnit{\state}{state}
\DeclareSIUnit{\bit}{bit}
\DeclareSIUnit{\mb}{MB}
\DeclareSIUnit{\mbit}{Mbit}
\DeclareSIUnit{\mb}{MB}

\newcommand{\cbox}[1]{{\color{#1}{$\blacksquare$}}}

%% file: main.bbl
\begin{thebibliography}{10}

\bibitem{efffuzz}
Efficient fuzzing guide.
\newblock
  \url{https://chromium.googlesource.com/chromium/src/+/main/testing/libfuzzer/efficient_fuzzing.md}.
\newblock Accessed Jun 6, 2024.

\bibitem{honggfuzz}
Honggfuzz.
\newblock \url{https://honggfuzz.dev/}.
\newblock Accessed Sep 15, 2024.

\bibitem{libfuzzer}
{libFuzzer} – a library for coverage-guided fuzz testing.
\newblock \url{https://llvm.org/docs/LibFuzzer.html}.
\newblock Accessed Feb 16, 2024.

\bibitem{murmurhash3}
Murmurhash3.
\newblock
  \url{https://blog.teamleadnet.com/2012/08/murmurhash3-ultra-fast-hash-algorithm.html}.
\newblock Accessed Sep 15, 2024.

\bibitem{google-ai}
Scaling security with ai: from detection to solution.
\newblock
  https://security.googleblog.com/2024/01/scaling-security-with-ai-from-detection.html.
\newblock Accessed Feb 16, 2024.

\bibitem{nikolaos2022longvuln}
Nikolaos Alexopoulos, Manuel Brack, Jan~Philipp Wagner, Tim Grube, and Max
  M{\"u}hlh{\"a}user.
\newblock How long do vulnerabilities live in the code? a {Large-Scale}
  empirical measurement study on {FOSS} vulnerability lifetimes.
\newblock In {\em 31st USENIX Security Symposium (USENIX Security 22)}, 2022.

\bibitem{aschermann2019redqueen}
Cornelius Aschermann, Sergej Schumilo, Tim Blazytko, Robert Gawlik, and
  Thorsten Holz.
\newblock {REDQUEEN:} fuzzing with input-to-state correspondence.
\newblock In {\em 26th Annual Network and Distributed System Security Symposium
  (NDSS)}, 2019.

\bibitem{bohme2021residual}
Marcel B\"{o}hme, Danushka Liyanage, and Valentin W\"{u}stholz.
\newblock Estimating residual risk in greybox fuzzing.
\newblock In {\em Proceedings of the 29th ACM Joint Meeting on European
  Software Engineering Conference and Symposium on the Foundations of Software
  Engineering}, 2021.

\bibitem{bohme2022reliability}
Marcel B\"{o}hme, L\'{a}szl\'{o} Szekeres, and Jonathan Metzman.
\newblock On the reliability of coverage-based fuzzer benchmarking.
\newblock In {\em Proceedings of the 44th International Conference on Software
  Engineering}, 2022.

\bibitem{chen2020savior}
Y.~Chen, P.~Li, J.~Xu, S.~Guo, R.~Zhou, Y.~Zhang, T.~Wei, and L.~Lu.
\newblock Savior: Towards bug-driven hybrid testing.
\newblock In {\em 2020 IEEE Symposium on Security and Privacy (SP)}, 2020.

\bibitem{chen2021polyglot}
Yongheng Chen, Rui Zhong, Hong Hu, Hangfan Zhang, Yupeng Yang, Dinghao Wu, and
  Wenke Lee.
\newblock One engine to fuzz ’em all: Generic language processor testing with
  semantic validation.
\newblock In {\em 2021 IEEE Symposium on Security and Privacy (SP)}, 2021.

\bibitem{deng2023nestfuzz}
Peng Deng, Zhemin Yang, Lei Zhang, Guangliang Yang, Wenzheng Hong, Yuan Zhang,
  and Min Yang.
\newblock Nestfuzz: Enhancing fuzzing with comprehensive understanding of input
  processing logic.
\newblock In {\em Proceedings of the 2023 ACM SIGSAC Conference on Computer and
  Communications Security}, 2023.

\bibitem{ubsan}
LLVM Developers.
\newblock Undefined behavior sanitizer, 2017.
\newblock \url{https://clang.llvm.org/docs/UndefinedBehaviorSanitizer.html}.

\bibitem{fioraldi2020aflpp}
Andrea Fioraldi, Dominik Maier, Heiko Ei{\ss}feldt, and Marc Heuse.
\newblock {AFL++} : Combining incremental steps of fuzzing research.
\newblock In {\em 14th USENIX Workshop on Offensive Technologies (WOOT 20)},
  2020.

\bibitem{gan2020greyone}
Shuitao Gan, Chao Zhang, Peng Chen, Bodong Zhao, Xiaojun Qin, Dong Wu, and
  Zuoning Chen.
\newblock {GREYONE}: Data flow sensitive fuzzing.
\newblock In {\em 29th USENIX Security Symposium (USENIX Security 20)}, 2020.

\bibitem{gan2018collafl}
Shuitao Gan, Chao Zhang, Xiaojun Qin, Xuwen Tu, Kang Li, Zhongyu Pei, and
  Zuoning Chen.
\newblock {CollAFL}: Path sensitive fuzzing.
\newblock In {\em 2018 IEEE Symposium on Security and Privacy (SP)}, 2018.

\bibitem{kasan}
Google.
\newblock The kernel address sanitizer, 2016.
\newblock \url{https://www.kernel.org/doc/html/latest/dev-tools/kasan.html}.

\bibitem{ktsan}
Google.
\newblock The kernel thread sanitizer, 2016.
\newblock \url{https://github.com/google/ktsan}.

\bibitem{kubsan}
Google.
\newblock The kernel undefined behavior sanitizer, 2016.
\newblock \url{https://www.kernel.org/doc/html/latest/dev-tools/ubsan.html}.

\bibitem{gross2023fuzzilli}
Samuel Gro{\ss}, Simon Koch, Lukas Bernhard, Thorsten Holz, and Martin Johns.
\newblock {FUZZILLI:} fuzzing for javascript {JIT} compiler vulnerabilities.
\newblock In {\em 30th Annual Network and Distributed System Security Symposium
  (NDSS)}, 2023.

\bibitem{lee2024syzrisk}
Solmaz Salimi Byoungyoung~Lee Gwangmu~Lee, Duo~Xu and Mathias Payer.
\newblock {SyzRisk}: A change-pattern-based continuous kernel regression
  fuzzer.
\newblock In {\em ACM ASIA Conference on Computer and Communications Security
  (ASIA CCS ’24)}, 2024.

\bibitem{harmouch2017cardinality}
Hazar Harmouch and Felix Naumann.
\newblock Cardinality estimation: An experimental survey.
\newblock {\em Proceedings of the VLDB Endowment}, 2017.

\bibitem{herrera2021seedsel}
Adrian Herrera, Hendra Gunadi, Shane Magrath, Michael Norrish, Mathias Payer,
  and Antony~L. Hosking.
\newblock Seed selection for successful fuzzing.
\newblock In {\em Proceedings of the 30th ACM SIGSOFT International Symposium
  on Software Testing and Analysis}, 2021.

\bibitem{kim2019hydra}
Seulbae Kim, Meng Xu, Sanidhya Kashyap, Jungyeon Yoon, Wen Xu, and Taesoo Kim.
\newblock Finding semantic bugs in file systems with an extensible fuzzing
  framework.
\newblock In {\em Proceedings of the 27th ACM Symposium on Operating Systems
  Principles}, 2019.

\bibitem{klees2018eval}
George Klees, Andrew Ruef, Benji Cooper, Shiyi Wei, and Michael Hicks.
\newblock Evaluating fuzz testing.
\newblock In {\em Proceedings of the 2018 ACM SIGSAC Conference on Computer and
  Communications Security}, 2018.

\bibitem{klees2018evaluating}
George Klees, Andrew Ruef, Benji Cooper, Shiyi Wei, and Michael Hicks.
\newblock Evaluating fuzz testing.
\newblock In {\em Proceedings of the ACM SIGSAC Conference on Computer and
  Communications Security (CCS)}, pages 2123--2138. ACM, 2018.

\bibitem{kong2024sand}
Ziqiao Kong, Shaohua Li, Heqing Huang, and Zhendong Su.
\newblock {SAND}: Decoupling sanitization from fuzzing for low overhead.
\newblock In {\em 2024 IEEE Symposium on Security and Privacy (SP)}, 2024.
\newblock Preprint.

\bibitem{li2021unifuzz}
Yuwei Li, Shouling Ji, Yuan Chen, Sizhuang Liang, Wei-Han Lee, Yueyao Chen,
  Chenyang Lyu, Chunming Wu, Raheem Beyah, Peng Cheng, Kangjie Lu, and Ting
  Wang.
\newblock {UNIFUZZ}: A holistic and pragmatic {Metrics-Driven} platform for
  evaluating fuzzers.
\newblock In {\em 30th USENIX Security Symposium (USENIX Security 21)}, 2021.

\bibitem{liu2023videzzo}
Qiang Liu, Flavio Toffalini, Yajin Zhou, and Mathias Payer.
\newblock Videzzo: Dependency-aware virtual device fuzzing.
\newblock In {\em 2023 IEEE Symposium on Security and Privacy (SP)}, 2023.

\bibitem{liu2023dsfuzz}
Yinxi Liu and Wei Meng.
\newblock Dsfuzz: Detecting deep state bugs with dependent state exploration.
\newblock In {\em Proceedings of the 2023 ACM SIGSAC Conference on Computer and
  Communications Security}, 2023.

\bibitem{metzman2021fuzzbench}
Jonathan Metzman, László Szekeres, Laurent Maurice~Romain Simon,
  Read~Trevelin Sprabery, and Abhishek Arya.
\newblock Fuzzbench: An open fuzzer benchmarking platform and service.
\newblock In {\em Proceedings of the 29th ACM Joint Meeting on European
  Software Engineering Conference and Symposium on the Foundations of Software
  Engineering}, 2021.

\bibitem{myung2022mundofuzz}
Cheolwoo Myung, Gwangmu Lee, and Byoungyoung Lee.
\newblock {MundoFuzz}: Hypervisor fuzzing with statistical coverage testing and
  grammar inference.
\newblock In {\em 31st USENIX Security Symposium (USENIX Security 22)}, 2022.

\bibitem{osterlund2020parmesan}
Sebastian \"{O}sterlund, Kaveh Razavi, Herbert Bos, and Cristiano Giuffrida.
\newblock Parmesan: Sanitizer-guided greybox fuzzing.
\newblock In {\em Proceedings of the 29th USENIX Conference on Security
  Symposium}, SEC'20, USA, 2020. USENIX Association.

\bibitem{papapetrou2010cardinality}
Odysseas Papapetrou, Wolf Siberski, and Wolfgang Nejdl.
\newblock Cardinality estimation and dynamic length adaptation for bloom
  filters.
\newblock {\em Distributed and Parallel Databases}, 2010.

\bibitem{rawat2017vuzzer}
Sanjay Rawat, Vivek Jain, Ashish Kumar, Lucian Cojocar, Cristiano Giuffrida,
  and Herbert Bos.
\newblock Vuzzer: Application-aware evolutionary fuzzing.
\newblock In {\em Proceedings of the Network and Distributed System Security
  Symposium (NDSS)}, 2017.

\bibitem{sergej2021nyx}
Sergej Schumilo, Cornelius Aschermann, Ali Abbasi, Simon W{\"o}r-ner, and
  Thorsten Holz.
\newblock Nyx: Greybox hypervisor fuzzing using fast snapshots and affine
  types.
\newblock In {\em 30th USENIX Security Symposium (USENIX Security 21)}, 2021.

\bibitem{schumilo2020hypercube}
Sergej Schumilo, Cornelius Aschermann, Ali Abbasi, Simon W{\"o}rner, and
  Thorsten Holz.
\newblock {HYPER-CUBE}: High-dimensional hypervisor fuzzing.
\newblock In {\em 27th Annual Network and Distributed System Security Symposium
  (NDSS)}, 2020.

\bibitem{serebryany2012asan}
Konstantin Serebryany, Derek Bruening, Alexander Potapenko, and Dmitriy Vyukov.
\newblock Addresssanitizer: A fast address sanity checker.
\newblock In {\em Proceedings of the USENIX Annual Technical Conference}, pages
  309--318, 2012.

\bibitem{song2020agamotto}
Dokyung Song, Felicitas Hetzelt, Jonghwan Kim, Brent~ByungHoon Kang,
  Jean-Pierre Seifert, and Michael Franz.
\newblock Agamotto: Accelerating kernel driver fuzzing with lightweight virtual
  machine checkpoints.
\newblock In {\em 29th USENIX Security Symposium (USENIX Security 20)}, 2020.

\bibitem{srivastava2021gramatron}
Prashast Srivastava and Mathias Payer.
\newblock Gramatron: effective grammar-aware fuzzing.
\newblock In {\em Proceedings of the 30th ACM SIGSOFT International Symposium
  on Software Testing and Analysis}, 2021.

\bibitem{stepanov2015msan}
Evgeniy Stepanov and Konstantin Serebryany.
\newblock {MemorySanitizer}: fast detector of uninitialized memory use in c++.
\newblock In {\em Proceedings of the 13th Annual IEEE/ACM International
  Symposium on Code Generation and Optimization}, 2015.

\bibitem{tarkoma2012bloom}
Sasu Tarkoma, Christian~Esteve Rothenberg, and Eemil Lagerspetz.
\newblock Theory and practice of bloom filters for distributed systems.
\newblock {\em IEEE Communications Surveys \& Tutorials}, 2012.

\bibitem{wang2019impactcov}
Jinghan Wang, Yue Duan, Wei Song, Heng Yin, and Chengyu Song.
\newblock Be sensitive and collaborative: Analyzing impact of coverage metrics
  in greybox fuzzing.
\newblock In {\em 22nd International Symposium on Research in Attacks,
  Intrusions and Defenses (RAID 2019)}, 2019.

\bibitem{wang2023fuzzjit}
Junjie Wang, Zhiyi Zhang, Shuang Liu, Xiaoning Du, and Junjie Chen.
\newblock {FuzzJIT}: {Oracle-Enhanced} fuzzing for {JavaScript} engine {JIT}
  compiler.
\newblock In {\em 32nd USENIX Security Symposium (USENIX Security 23)}, 2023.

\bibitem{wang2020notallcov}
Yanhao Wang, Xiangkun Jia, Yuwei Liu, Kyle Zeng, Tiffany Bao, Dinghao Wu, and
  Purui Su.
\newblock Not all coverage measurements are equal: Fuzzing by coverage
  accounting for input prioritization.
\newblock In {\em 27th Annual Network and Distributed System Security Symposium
  (NDSS)}, 2020.

\bibitem{wu2022strategy}
Mingyuan Wu, Ling Jiang, Jiahong Xiang, Yanwei Huang, Heming Cui, Lingming
  Zhang, and Yuqun Zhang.
\newblock One fuzzing strategy to rule them all.
\newblock In {\em 2022 IEEE/ACM 44th International Conference on Software
  Engineering (ICSE)}, 2022.

\bibitem{wu2023jitfuzz}
Mingyuan Wu, Minghai Lu, Heming Cui, Junjie Chen, Yuqun Zhang, and Lingming
  Zhang.
\newblock Jitfuzz: Coverage-guided fuzzing for jvm just-in-time compilers.
\newblock In {\em Proceedings of the 45th International Conference on Software
  Engineering}, 2023.

\bibitem{yan2020pathafl}
Shengbo Yan, Chenlu Wu, Hang Li, Wei Shao, and Chunfu Jia.
\newblock {PathAFL}: Path-coverage assisted fuzzing.
\newblock In {\em Proceedings of the 15th ACM Asia Conference on Computer and
  Communications Security}, 2020.

\bibitem{yun2016apisan}
Insu Yun, Changwoo Min, Xujie Si, Yeongjin Jang, Taesoo Kim, and Mayur Naik.
\newblock {APISan}: Sanitizing {API} usages through semantic {Cross-Checking}.
\newblock In {\em 25th USENIX Security Symposium (USENIX Security 16)}, 2016.

\bibitem{afl}
M.~Zalewski.
\newblock American fuzzy lop, 2017.
\newblock \url{http://lcamtuf.coredump.cx/afl/technical_details.txt}.

\bibitem{zhai2022ndss}
Yizhuo Zhai, Yu~Hao, Zheng Zhang, Weiteng Chen, Guorern Li, Zhiyun Qian,
  Chengyu Song, Manu Sridharan, Srikanth~V. Krishnamurthy, Trent Jaeger, and
  Paul Yu.
\newblock {Progressive Scrutiny: Incremental Detection of UBI bugs in the Linux
  Kernel}.
\newblock In {\em Proceedings of the 2020 ISOC Network and Distributed Systems
  Security Symposium (NDSS)}, 2022.

\bibitem{zheng2023fishfuzz}
Han Zheng, Jiayuan Zhang, Yuhang Huang, Zezhong Ren, He~Wang, Chunjie Cao,
  Yuqing Zhang, Flavio Toffalini, and Mathias Payer.
\newblock {FISHFUZZ}: Catch deeper bugs by throwing larger nets.
\newblock In {\em 32nd USENIX Security Symposium (USENIX Security 23)}, 2023.

\bibitem{aflchurn}
Xiaogang Zhu and Marcel B{\"o}hme.
\newblock Regression greybox fuzzing.
\newblock In {\em Proceedings of the 2021 ACM SIGSAC Conference on Computer and
  Communications Security}, pages 2169--2182, 2021.

\end{thebibliography}
